\documentclass{article}

  \usepackage[preprint]{neurips_2026}

\usepackage[utf8]{inputenc} 
\usepackage[T1]{fontenc}    
\usepackage{url}            
\usepackage{booktabs}       
\usepackage{amsfonts}       
\usepackage{nicefrac}       
\usepackage{microtype}      
\usepackage{xcolor}         
\usepackage{amsmath}
\usepackage{subcaption} 
\usepackage{comment} 
\usepackage{algorithm}
\usepackage{algpseudocode}
\usepackage{rotating}

\usepackage{natbib}
\setcitestyle{authoryear,round,citesep={;},aysep={,},yysep={;}}
\definecolor{mydarkblue}{rgb}{0,0.08,0.45}
\usepackage[colorlinks=true, linkcolor=mydarkblue, citecolor=mydarkblue, urlcolor=mydarkblue]{hyperref}
\usepackage{graphicx} 

\title{Bayesian Nonparametric Clustering to Support Medical Decision-Making: A Variational Inference Approach}

\author{%
  Inga Huld~Ármann \\
  Department of Mathematics\\
  Imperial College London\\
  London, UK \\
  \texttt{inga.armann23@imperial.ac.uk} \\
   \And
   Ioanna Papatsouma \\
   Department of Mathematics\\
  Imperial College London\\
  London, UK \\
   \texttt{i.papatsouma@imperial.ac.uk} \\
   \AND
   Marina Evangelou \\
   Department of Mathematics\\
  Imperial College London\\
  London, UK \\
\texttt{m.evangelou@imperial.ac.uk} \\
}

\begin{document}

\maketitle

\begin{abstract}

Medical decision-making increasingly requires rapid and reliable assignment of patients to disease subtypes, as many diseases are no longer treated as single entities. For example, cancer patients may be stratified into aggressive and non-aggressive subtypes, with different treatment strategies for each group. We propose a Bayesian nonparametric approach based on a Dirichlet process mixture model for clustering individuals into disease subtypes. We implement a coordinate ascent variational inference algorithm, yielding an effective and computationally efficient alternative to Markov chain Monte Carlo (MCMC), to support medical decision-making. In synthetic experiments, we demonstrate that the proposed approach accurately assigns observations to their ground-truth clusters, achieving strong performance across evaluation metrics, such as homogeneity and completeness. Additionally, we illustrate the proposed approach achieves a substantial improvement in computational cost compared to MCMC, without sacrificing accuracy that would lead to the increased risk of misdiagnosis.

\end{abstract}

\section{Introduction}\label{sec:introduction}

Decision-making in biological and healthcare settings is inherently complex and requires models that can capture uncertainty, heterogeneity, and high-dimensional structure in data. Patient populations are rarely homogeneous, meaningful subgroups often emerge only through careful analysis of underlying patterns and incorrect decisions can potentially lead to severe consequences. A medical decision-making workflow requires rapid and reliable assignment of individuals into disease subtypes. Our solution to this challenge is the implementation of a variational inference (VI) approximation for a Bayesian nonparametric model as a computationally efficient and reliable tool to support medical decision-making. 

\begin{figure}[!h]
  \centering
\includegraphics[width=0.95\linewidth]{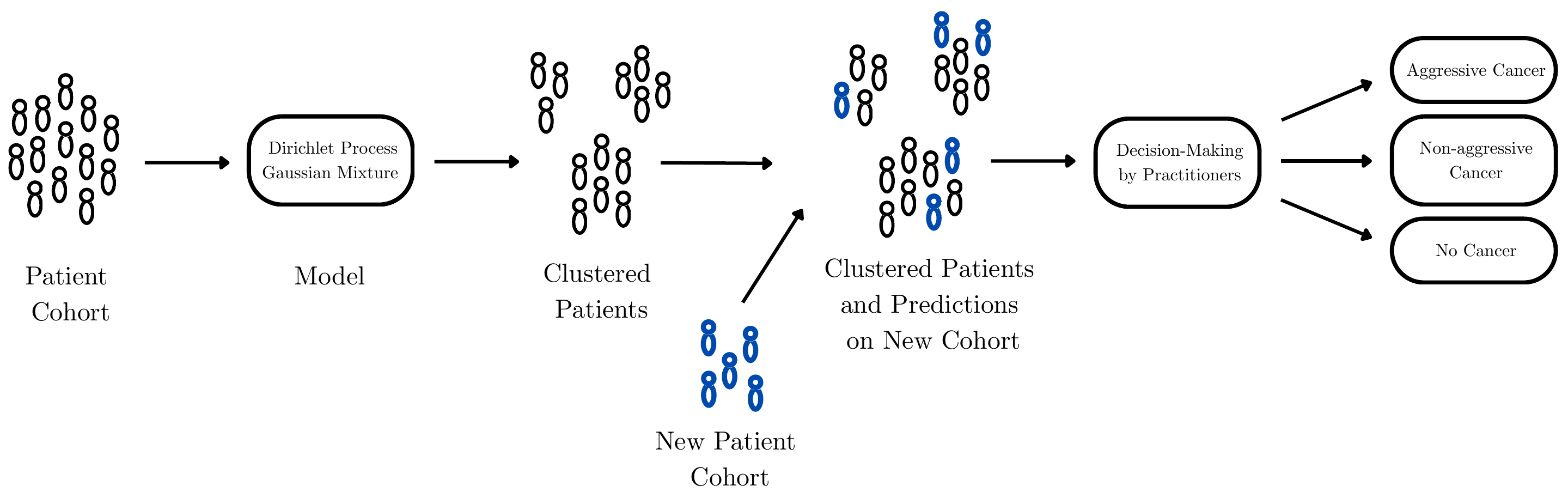}
  \caption{Illustration of a medical decision-making workflow for healthcare practitioners,
   supported by clustering with Dirichlet process Gaussian mixture models.}
  \label{fig:keyfigure}
\end{figure}

 As illustrated in Figure~\ref{fig:keyfigure}, the proposed approach can be applied to a population of patients to infer the underlying structure of the data and recover groups within the population. Subsequently, as new patients are recruited, the proposed model can predict their allocated clusters. This can enable further insights into patient cohorts and hence provide a tool to support medical decision-making. 
 
 From a decision-making perspective, there is currently limited evidence on how approximations, like the ones offered by VI, can specifically inform algorithmic choices in clinical practice. We explore the performance of the proposed approach through a numerical study that provides a controlled framework. The presented synthetic experiments are motivated by healthcare scenarios, with  systematic variation in parameters to reflect settings encountered in healthcare applications.

 Variational inference has often been demonstrated to provide fast and accurate solutions to various Bayesian problems, including clustering. But in the specific application to medical decision-making, more care is required than solely relying on a single clustering evaluation metric, providing an incomplete picture of the performance. We propose the use of metrics that capture multiple aspects of clustering quality, including completeness, the extent to which members of a true group are assigned to the same cluster, and homogeneity, the purity of inferred clusters.

Clustering through Bayesian nonparametric models, including the most popular one, Dirichlet process mixture model (DPMM), allows for probabilistic assignments and more interpretable results. A key advantage of DPMM lies in additional flexibility by allowing for an infinite number of clusters, inferring the number of clusters directly from the data, rather than requiring it to be fixed in advance. In this work, we focus on a mixtures of Gaussian distributions, introduced by \citet{escobar1995bayesian}, which we abbreviate as DPGMM. By implementing a VI approximation for DPGMM we obtain a scalable and efficient inference of the model enabling robust clustering in high-dimensional and noisy healthcare data to support rapid medical decision-making workflow.

Inference in Bayesian models has traditionally relied on Markov Chain Monte Carlo (MCMC) methods, which provide asymptotically exact approximations to posterior distributions. Despite their theoretical appeal, MCMC methods are often computationally expensive and can be impractical for large-scale or time-sensitive applications. VI offers an alternative by reframing inference as an optimisation problem, potentially trading accuracy for substantial gains in computational efficiency \citep{blei2017variational}. One of the key questions addressed in the presented work is whether the computational efficiency of the VI approximation comes with a cost on clustering performance and prediction accuracy. Through synthetic data experiments, we show that the VI implementation offers outstanding clustering quality while achieving orders of magnitude shorter computational time compared to MCMC. The results of this study therefore help clarify their relative suitability in applied settings.

Our main contribution is to lay the foundation for integrating a VI implementation for DPGMM into a medical decision-making workflow as illustrated in Figure \ref{fig:keyfigure}. In doing so, we bridge a gap between scalable Bayesian nonparametric methods and their practical use in clinical decision support. This supports more reliable patient subtyping for informed treatment planning in data-driven healthcare settings, saving time and resources.

\section{Background} \label{sec:background}

This section presents the general Dirichlet process mixture model for Bayesian clustering and the variational inference approach of \citet{blei2006variational}. The specific model used in this paper and the corresponding variational inference framework is introduced in Section \ref{sec:model}. 

Clustering provides a natural framework for uncovering latent structure in biological data, enabling the identification of subpopulations that may correspond to disease subtypes, treatment responses, or risk profiles. 
Bayesian approaches to clustering extend this framework by explicitly modelling uncertainty, allowing for probabilistic assignments and more interpretable results. The Dirichlet process mixture model (DPMM) builds on the Dirichlet process of \citet{ferguson1973bayesian} and was further extended to mixture models in \citet{escobar1995bayesian}. DPMMs have been successfully applied in various biological settings, including clustering of tissue samples from MRI images in \citet{da2007dirichlet}, pancreatic cancer gene expression data in \citet{hossain2021discovering} and Parkinson's disease patient symptom data \citet{white2012dirichlet}. An overview of Dirichlet processes is provided by \citet{teh2010dirichlet}. 

\subsection{Dirichlet Process Mixture Models}

Given data observations $X = \{x_i\}_{i=1}^N$, with each data observation $x_i$ containing $P$ features, $x_i\in \mathbb{R}^P$, and a set of latent parameters $\{\theta_i\}_{i=1}^N$, we model each data observation $x_i$ with the distribution $F(\theta_i)$. The Dirichlet process mixture model is denoted as follows,
 
\begin{align}\label{eq:DPMM_1}
    x_i | \theta_i &\sim F(\theta_i) \nonumber \\
    \theta_i | G &\sim G \nonumber \\ 
    G|\alpha, H &\sim DP(\alpha, H)
\end{align}

Where DP denotes the Dirichlet process distribution with concentration parameters $\alpha$ and base distribution $H$. 
Due to the discrete nature of the DP, draws $\theta_i$ from the distribution $G$ are repeated, as they do not need to be unique. Defining each unique recurrence of $\theta_i$ as $\theta_i^*$, we obtain a clustering property where each $\theta^*_i$ defines a cluster. 
When implementing DPs in practice the stick-breaking construction is a common form of the DP \citep{sethuraman1994constructive}. It provides a simple and intuative approach to clustering with a mixture model. In the stick-breaking construction of a DP, we specify the following, 

\begin{align*}
    \beta_k \sim \text{Beta}(1,\alpha) &\qquad \theta^*_k\sim H\\
    \pi_k = \beta_k \Pi^{k-1}_{l=1}(1-\beta_l) &\qquad G=\sum^\infty_{k=1}\pi_k \delta_{\theta^*_k},
\end{align*}

then $G\sim DP(\alpha, H)$. For convenience, the stick-breaking construction may be denoted as $\pi \sim GEM(\alpha)$, where GEM denotes the Griffiths-Engen-McCloskey distribution \citep{pitman1996combinatorial}.

Denote $z_i$ as a cluster assignment variable, taking the value $k$ with probability $\pi_k$, with $k = 1,2,...$. Then the latent parameters may be written as the unique cluster parameters with the following notation $\theta_i = \theta_{z_i}^*$, hence $F(\theta_k^*)$ is the distribution of the data observations for cluster $k$. Model~\ref{eq:DPMM_1} may then be rewritten as the following,

\begin{align}\label{eq:DPMM_2}
    \pi | \alpha \sim \text{GEM}(\alpha) &\qquad \theta^*_k | H \sim H \nonumber \\
    z_i | \pi \sim \text{Categorical}(\pi) &\qquad x_i | z_i, \{\theta^*_k\} \sim F(\theta^*_{z_i}).
\end{align}

 As noted in \cite{teh2010dirichlet}, the DPMM is an infinite mixture model with the appealing property that only a few clusters are inferred in practice. This stems from the property that $\pi_k$ decreases exponentially quickly. Thus, the number of clusters does not need to be defined beforehand, giving DPMMs an advantage over finite mixture models. 

A variety of MCMC methods have been developed for DPMMs including blocked Gibbs sampling in \cite{ishwaran2001gibbs}, collapsed Gibbs sampling in \cite{maceachern1994estimating}, Gibbs samplers for DPMMs in \cite{neal2000markov} and a split-merge MCMC in \cite{jain2004split}. 

\subsection{Variational Inference}

As an alternative to MCMC, variational inference provides an optimisation approach to approximate the posterior distribution of interest. Given a variational family of distributions, $\mathcal{Q}$, variational inference approximates the posterior by a distribution $q^*\in \mathcal{Q}$. The optimal $q^*$ is found by miniminising the Kullback-Leibler divergence between the variational distribution $q$ and the exact posterior. 

For computing the Kullback-Leibler divergence, the logarithmic evidence is required, $\log p(\boldsymbol{x})$, which is often intractable. Using the fact that $KL(\cdot) \geq 0$, a lower bound is derived for the logarithmic evidence, known as the evidence lower bound (ELBO). Given the data observations $\boldsymbol{x}$ and latent parameters $\boldsymbol{w}$, the ELBO is the following, 

\begin{align}\label{eq:ELBO}
    \log p(\boldsymbol{x}) \geq E_q[\log p(\boldsymbol{w}, \boldsymbol{x})] - E_q[\log q(\boldsymbol{w})] = ELBO(q).
\end{align}

This reformulates the minimisation problem of the Kullback-Leibler divergence as a maximisation problem of the ELBO in Equation~\ref{eq:ELBO}.  

As in \citet{blei2006variational}, we choose the \textit{mean-field variational family} to maximise over and assume the optimal distribution for each latent parameter, $\boldsymbol{w}$, is a member of the exponential family. Futher, we truncate the variational distribution at a level $T$, such that $q(\beta_T = 1) = 1$, and for $k>T$, $\pi_k = 0$. We set the latent parameters as $\boldsymbol{w} = \{\boldsymbol{\beta}, \boldsymbol{\theta}^*, \boldsymbol{z}\}$. The corresponding fully factorised family of variational distributions for the mean-field variational inference for DPMM is the following,

\begin{align*}
    q(\boldsymbol{\beta}, \boldsymbol{\theta}^*, \boldsymbol{z}) = \prod^{T-1}_{t=1}q_{\rho_t}(\beta_t) \prod^{T}_{t=1}q_{\tau_{t}}(\theta^*_{t})
    \prod^{N}_{n=1}q_{\phi_n}(z_n),
\end{align*}

where $q_{\rho_k}(\beta_k)$ is a Beta distribution with the natural parameters $\{\rho_{k,1},\rho_{k,2}\}$, $q_{\tau_k}(\theta^*_k)$ is a exponential family distribution with natural parameters $\{\tau_{k,1},\tau_{k,2}\}$ and $q_{\phi_n}(z_n)$ is a categorical distribution with parameters $\{\phi_{n,1},...,\phi_{n,T}\}$. Further, in \citet{blei2006variational} they derive general update steps for DPMM with exponential family distributions for the mixture distribution with coordinate ascent  variational inferece (CAVI). The optimal distribution, $q^*_{\nu_j}$ for each latent variable $\boldsymbol{w}_j$ is the following, 

\begin{align}\label{eq:CAVI_general}
    q^*_{\nu_j}(w_j) &\propto \exp\left\{E_{-j}[\log p(w_j,\boldsymbol{w}_{-j}, \boldsymbol{x})]\right\}.
\end{align}

The optimal distribution from Equation~\ref{eq:CAVI_general} is then a coordinate update for CAVI. The variational parameters are iteratively updated using the corresponding coordinate steps and the ELBO computed, until the ELBO has converged. When the ELBO has converged the end result is the corresponding variational distribution with the final variational parameters. Further, the variational distribution may then be used as an approximation of the posterior for the model of interest. 

\section{Dirichlet Process Gaussian Mixture Model} \label{sec:model}

We now introduce the proposed model and derive the corresponding CAVI update steps. The proposed model will be used as a tool to support decision-making and is evaluated on synthetic data in Section~\ref{sec:numerical_studies}. Specifically, we consider a Dirichlet process Gaussian mixture model (DPGMM), in which the mixtures are Gaussian distributions with unknown means and variances. The base measure $H$ is set as a product measure with conjugate priors for the mixture distribution parameters, $\mu_k$ and $\sigma^2_k$. For data points $\{x_i\}_{i=1}^N$, the model is the following,

\begin{align*}
    x_i \mid z_i   &\sim \mathcal{N}(\mu_{z_i}, \sigma^2_{z_i} I)
&\qquad z_i \mid \pi  &\sim \text{Categorical}(\pi)
&\qquad \pi           &\sim \text{GEM}(\alpha) \\
\mu_k                 &\sim \mathcal{N}(\mu_0, \sigma_0^2 I)
&\qquad \sigma_k^2    &\sim \text{Inverse-Gamma}(b,c)
&\qquad \alpha        &\sim \text{Gamma}(s_1,s_2).
\end{align*}

The explicit form of the ELBO for the DPGMM is provided in Appendix~\ref{sec:elbo}. To maximise the ELBO from Equation~\ref{eq:ELBO} with respect to the variational parameters, we derive the coordinate ascent update steps by applying the optimal distribution in Equation~\ref{eq:CAVI_general}. The update steps take the following form,

\begin{enumerate}

  \item $q_{\phi_n}(z_n)$ is a Categorical distribution with parameters $(\phi_{n,1},...,\phi_{n,T})$,

\begin{align*}
    \phi_{n,k} &= \exp\biggl\{\psi(\rho_{k,1})-\psi(\rho_{k,1}+\rho_{k,2})+ \sum^{k-1}_{t=1}\psi(\rho_{t,2})-\psi(\rho_{t,2}+\rho_{t,1}) \\&+ P \log (2\pi)+P\left[\log (\tau^{\sigma^2}_{k,2}) + \psi(\tau^{\sigma^2}_{k,1})\right] - \frac{\tau^{\sigma^2}_{k,1}}{2\tau^{\sigma^2}_{k,2}}\left[||x_{n}-\tau^{\mu}_{k,1}||_2^2 + \text{trace}(\tau^{\mu}_{k,2})\right]
 \biggl\}.
\end{align*}

    \item $q_{\rho_k}(\beta_k) $ is a Beta distribution with natural parameters $(\rho_{k,1},\rho_{k,2})$,
 \begin{align*}
    \rho_{k,1} = 1+ \sum_{n=1}^{N} \phi_{n,k}, \qquad
    \rho_{k,2} = \frac{w_1}{w_2} + \sum_{n=1}^{N} \sum^T_{l=k+1}\phi_{n,l}.
\end{align*}

\item $q_{w}(\alpha)$ is a Gamma distribution with parameters $(s_1, s_2)$,

\begin{align*}
    w_1 = s_1 + T -1, \qquad
    w_2 = s_2 - \sum^{T-1}_{k=1}\left[ \psi(\rho_{k,2})- \psi(\rho_{k,1} + \rho_{k,2}) \right]
\end{align*}

   \item  $q_{\tau^{\mu}_{k}}(\mu^*_{k})$ is a Gaussian distribution with mean $\tau^{\mu}_{k,1}$ and covariance $\tau^{\mu}_{k,2}$, 

    \begin{align*}
        \tau^{\mu}_{k,1} = \frac{ (\sigma^2_0I)^{-1}\mu_0+\frac{\tau^{\sigma^2}_{k,1}}{\tau^{\sigma^2}_{k,2}}\sum_n\phi_{n,k}x_{n}}
        {(\sigma^2_0I)^{-1}  + \frac{\tau^{\sigma^2}_{k,1}}{\tau^{\sigma^2}_{k,2}}\sum_n\phi_{n,k}}, \qquad
        \tau^{\mu}_{k,2} = \frac{1}{(\sigma^2_0I)^{-1} + \frac{\tau^{\sigma^2}_{k,1}}{\tau^{\sigma^2}_{k,2}}\sum_n\phi_{n,k}}.
    \end{align*}
  
    \item  $q_{\tau^{\sigma^2}_{k}}(\sigma^{2^*}_{k})$ is a Inverse-Gamma distribution with shape $\tau^{\sigma^2}_{k,1}$ and scale $\tau^{\sigma^2}_{k,2}$,

    \begin{align*}
        \tau^{\sigma^2}_{k,1} = b + \frac{P}{2}\sum_n\phi_{n,k}, \qquad
        \tau^{\sigma^2}_{k,2} = c + \frac{1}{2} \sum_n \phi_{n,k}\left[||x_{n}-\tau^{\mu}_{k,1}||_2^2 + \text{trace}(\tau^{\mu}_{k,2})\right].
    \end{align*}

\end{enumerate}

These update steps define the CAVI algorithm, a computationally efficient procedure for approximating the posterior of the Bayesian nonparametric clustering model, DPGMM, implemented in Section~\ref{sec:numerical_studies}.

\section{Numerical Studies}\label{sec:numerical_studies}

In this section, we consider simulated data designed to reflect a range of clinically relevant scenarios. Because the ground-truth labels are known by construction, this enables a controlled evaluation of model performance. We first describe the data-generating process, where the synthetic data are created with incorporated noise and correlation to better capture characteristics observed in biomedical data. We then outline the performance metrics used to evaluate the proposed approach. Section~\ref{sec:data_gen_eval} concludes with a brief description of the implementation details where full details are provided in Appendix~\ref{sec:numerical_study_setup}. The findings from the experiments conducted are presented in Section \ref{sec:results} and additional findings are included in Appendix~\ref{sec:appendix_results}. 

\subsection{Data Generation and Evaluation}\label{sec:data_gen_eval}

\paragraph{Synthetic Data}
We created synthetic data that capture characteristics of real biomedical data. To achieve this we have generated data from a finite mixture of $K$ Gaussian distributions. The number of features is fixed across all presented experiments at $P= 200$, of which $P_{\text{noise}} = 50$ features are drawn from $\mathcal{N}(0,1)$ representing noisy features. The mixture component mean parameters of the Gaussian distributions are drawn from $\mu_k\sim\mathcal{N}(0,5^2)$ and the variance parameters from $\sigma^2_k \sim \text{Inverse-Gamma}(2,1)$, with $k = 1,...,K$. For each component, the same mean and variance parameter is used across features. The mixture proportions are drawn from $\text{Dirichlet}(10\cdot \boldsymbol{1}_k)$. An AR(1) correlation structure was introduced to the data to better reflect real biomedical data and additive Gaussian noise is introduced to each observation. In the experiments presented in the following sections, the sample size $N$ is varied between $\{50, 200, 500, 1000\}$ to assess how runtime scales with the number of observations. Further, the number of disease subtypes is varied between 2 and 11, where $K=2$ corresponds to no disease versus disease, $K=3$ corresponds to no disease, non-aggressive disease and aggressive disease, and $K>3$ corresponds to multiple disease subtypes. In the following experiments, the predictive performance of the proposed model is explored by splitting the data into training and test sets, where the test set is fixed at $N_{\text{test}} = 10$ samples to reflect a realistic clinical setting that involves a small number of new patients. 

\paragraph{Performance Evaluation}

The cluster assignments obtained are evaluated through several clustering evaluation measures, including homogeneity and completeness. Homogeneity measures whether each cluster contains only observations with the same ground truth label.  Completeness measures whether observations with the same ground truth label are assigned to the same cluster. Both metrics range from 0 to 1, with higher values indicating better satisfaction of the criteria \citep{rosenberg2007v}. We use completeness as a clinically relevant metric, as splitting observations from the same ground truth subtype across multiple inferred cluster assignments may increase risk of misdiagnosis when using the model to support clinical decision-making.  Two extrinsic measures are evaluated that compare the obtained clusters with external information (ground truth): the Adjusted Rand Index (ARI) \citep{hubert1985comparing} and the Normalised Mutual Information (NMI) score \citep{strehl2002cluster}. ARI ranges from 
-1 to 1 with 1 being a perfect match, 0 being a random clustering and negative values indicate agreement worse than random. On the other hand, NMI ranges from 0 to 1, where 1 denotes a perfect match between the clustering assignments and the known labels. These four measures are evaluated on both the training and the test sets across the different scenarios investigated. In addition to them, the intrinsic measure, silhouette score \citep{rousseeuw1987silhouettes} that measures how close the obtained clusters are to each other without using external information, is evaluated only on the training set.
We further report wall clock runtimes for all methods in seconds, for both inference and prediction. 

\paragraph{Method Implementation}
CAVI for the DPGMM is implemented using the \texttt{Python} code provided in \url{https://github.com/ingahuld/DPMM}. We run CAVI until the ELBO reaches convergence criteria of $10^{-5}$. 
We compare our CAVI implementation with MCMC using \texttt{PyMC} \citep{abril2023pymc} which employs the No-U-Turn sampler (NUTS) \citep{hoffman2014no}.

\subsection{Results} \label{sec:results}

\paragraph{VI approximation for DPGMM as a decision-making clustering tool}

 We firstly evaluate the performance of our VI approximation for DPGMM as a tool to support decision-making for disease subtyping. For both the simpler cases, $K = 2$ and $K = 3$, the VI approximation is able to recover the underlying structure of the training set and make accurate predictions on the test set (Table~\ref{tab:vi_alpha_prior_n200_k23}, Appendix Tables~\ref{tab:vi_alpha_prior_train} and ~\ref{tab:vi_alpha_prior_test}), across different tested $N$ values. 
 In some cases, the model inferred more clusters than in the ground truth, slightly reducing ARI, NMI, and completeness on the training set, but leaving test-set performance unaffected, with all three metrics equal to 1 (Appendix Tables~\ref{tab:vi_alpha_prior_train} and ~\ref{tab:vi_alpha_prior_test}).
 Despite this small number of cases, the proposed VI approximation recovered the underlying structure well overall, with perfect predictions on the test set, indicating that it performs well in these lower complexity settings. As the underlying structure of the data becomes more complex, with more subgroups, a drop in ARI, NMI, homogeneity is observed, while completeness is maintained at 1 for both the training and test sets (Appendix Figure~\ref{fig:vi_repeats_metrics_appendix}). In the following experiments, the values of $K = 2, 3, 7$ are explored, with the experiments comparing MCMC with VI concentrated on the more complex case, $K = 7$. 
 
\begin{table*}[!h]
\centering
\caption{{\bf VI approximation for DPGMM as a decision-making clustering tool}. VI test set clustering performance with a prior on $\alpha$, for $N=200$ and $K=2, 3$. The inferred number of clusters, inference and prediction runtime, and test set ARI, NMI, homogeneity, and completeness are reported. VI achieves perfect test set clustering quality in both settings. }
\label{tab:vi_alpha_prior_n200_k23}
\begin{tabular}{cccccccc}
\toprule
$K$ & Clust. &Time$_{\text{inf.}}$ (s) & Time$_{\text{pred.}}$ (s) & ARI$_{\text{test}}$ & NMI$_{\text{test}}$ & Homog.$_{\text{test}}$ & Compl.$_{\text{test}}$ \\
\midrule
2 & 3 & 2.11 & 0.00 & 1.0000 & 1.0000 & 1.0000 & 1.0000 \\
3 & 3 & 0.73 & 0.00 & 1.0000 & 1.0000 & 1.0000 & 1.0000 \\ 
\bottomrule
\end{tabular}
\end{table*}
\paragraph{Effect of the concentration parameter $\alpha$}

\begin{figure}[!h]
  \centering
      \includegraphics[width=\linewidth]{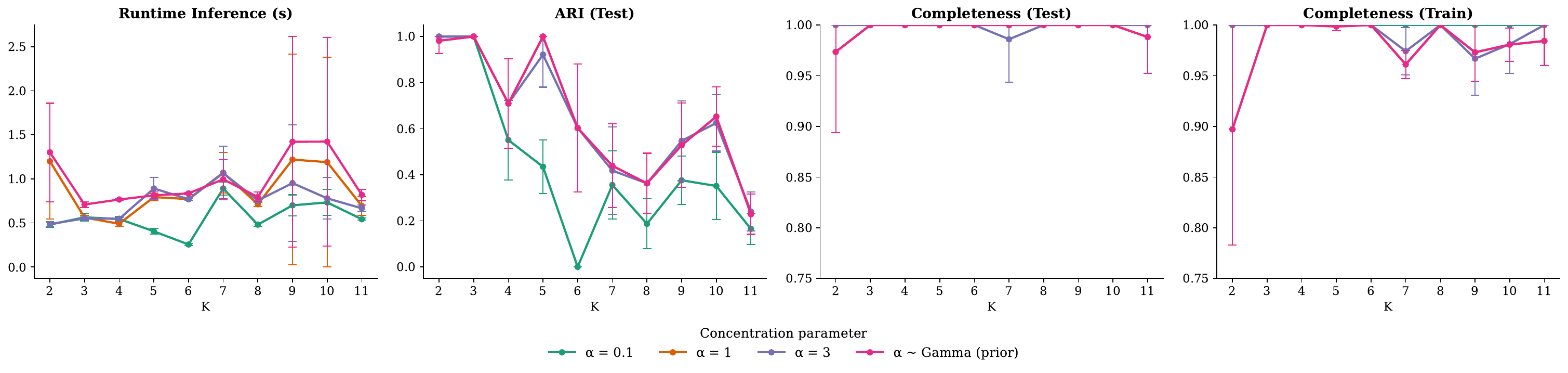}
  \caption{{\bf Effect of the concentration parameter $\alpha$}. VI clustering performance for $K \in \{2,3,...,10,11\}$ using fixed values of $\alpha \in\{0.1,1,3\}$ and a $\text{Gamma}(1,1)$ prior. The mean, and standard deviation over 10 train/test splits are illustrated for inference runtime, test set ARI, training and test sets completeness. The models corresponding to $\alpha =1$, $\alpha = 3$ and $\alpha \sim \text{Gamma}(1,1)$ exhibit similar performance whereas the model with $\alpha = 0.1$ performs comparatively worse.}
  \label{fig:vi_repeats_metrics}
\end{figure}

The effect of the concentration parameter $\alpha$ is investigated by applying variational inference with fixed values, $\alpha \in \{0.1, 1, 3\}$ and with a prior on alpha such that $\alpha \sim \text{Gamma}(1,1)$ across $K \in \{2, 3, .., 10, 11\}$ over 10 independent random train/test splits, with $N = 200$ fixed. Figure~\ref{fig:vi_repeats_metrics} illustrates how larger values of $\alpha$ perform better compared to smaller values, with the prior on $\alpha$ to show the most robust behaviour across the different grouping structures explored. These observations are made across all computed clustering evaluation metrics (Appendix Figure~\ref{fig:vi_repeats_metrics_appendix}). For the remaining experiments, a prior $\text{Gamma}(1,1)$ is assumed for $\alpha$ as it provides the most flexible implementation of the proposed model. 

\paragraph{Effect of Training Size} 
We evaluate the effect of training set size on the clustering performance of the VI approximation. In biomedical studies, training-set sizes can range from very small to several thousand observations, so any implemented methods must be both efficient and effective across the different settings.

\begin{figure}[!h]
  \centering
      \includegraphics[width=\linewidth]{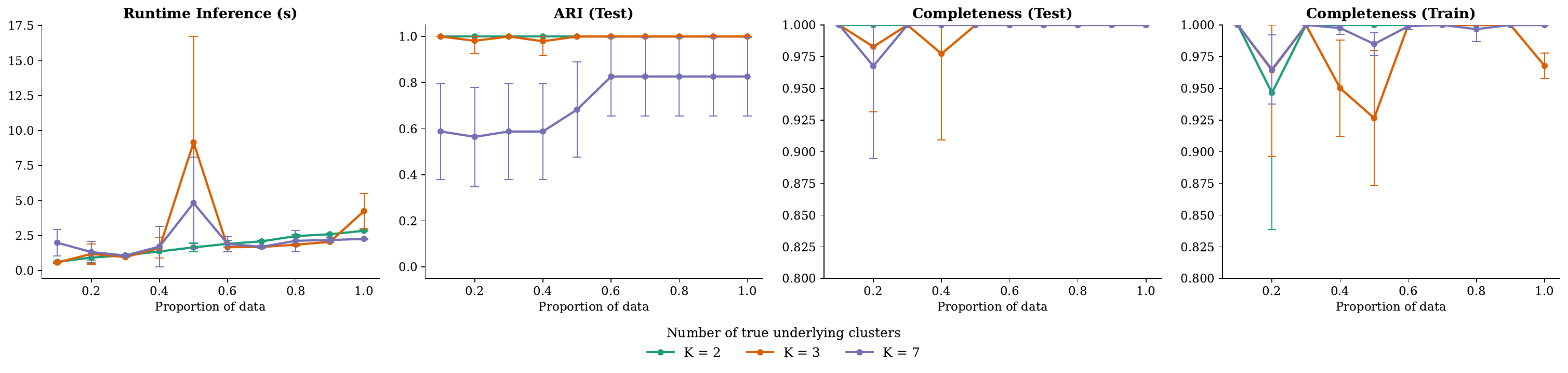}
  \caption{\textbf{Effect of training size}. VI clustering performance as the number of training observations increases, with $N_{\text{train}} = 1000p$ for $p\in\{0.1,0.2,...,0.9,1.0\}$. The mean, and standard deviation over 10 train/test splits are illustrated for inference runtime, test set ARI, training and test set completeness. Clustering performance is generally high across $p$ for $K = 2$ and $K = 3$. For $K = 7$ it is lower for smaller $p$ and improves until stabilising for $p$ values greater than $0.6$. As expected the inference runtime increases with $N_{\text{train}}$ but broadly remains below 10 seconds.}
  \label{fig:vi_repeats_increaseN}
\end{figure}

We use the simulated dataset created for $N = 1000$ to subsample new training subsets of size $N_{\text{train}} = 1000p$, where $p\in\{0.1,0.2,...,0.9,1.0\}$. As previously, the test set for all these different sets is the same for all values of $p$ and set as $N_{\text{test}}=10$ to better reflect real case scenarios. Figure~\ref{fig:vi_repeats_increaseN} illustrates that the predictive performance improves as the number of training set increases, as reflected by both the ARI and completeness metrics. Similar results are observed for the additional performance metrics (Appendix Figure~\ref{fig:vi_repeats_metrics_increaseN_appendix}). Performance is broadly consistent across $p$ for $K = 2$ and $K = 3$, although a few outliers appear in completeness around $p = 0.2$ for $K = 2$ and $p = 0.4$ and $p = 0.6$ for $K = 3$. These same values also show outliers in inference runtime, indicating slower convergence. For $K = 7$, ARI improves as $p$ increases before stabilising after $p = 0.6$. As expected, the main trend is that inference runtime increases with $N_{\text{train}}$ but still remains generally below 5 seconds which is an invaluable characteristic for a decision-making tool.  

\paragraph{Comparison with MCMC} 
The performance of our proposed VI approximation is compared with its counterpart competitor, MCMC. MCMC offers an exact solution to the posterior inference of DPGMM, whereas VI is an approximation. By comparing these two solutions, we aim to evaluate the benefits of the proposed approximation against the risk of misdiagnosis. 

As VI showed excellent performance for less complex cases, we demonstrate the performance of VI and MCMC for $K = 7$ for different values of $N$. As illustrated in Table~\ref{tab:k7_alpha_comparison_compact}, VI has significantly faster runtime compared to MCMC and perfect or near to perfect completeness of clusters. VI outperforms MCMC in terms of ARI on training and test set for $N = 1000$ and $N = 50$ whilst MCMC outperforms for $N = 200$ and $N = 500$. Similar results were observed for the other performance metrics as well. In terms of the completeness matric, the VI approximation is found to outperform MCMC, on both the training and test sets (Appendix Tables~\ref{tab:k7_alpha_comparison_train} and~\ref{tab:k7_alpha_comparison_test}). 

\begin{table*}[!h]
\centering
\small
\caption{{\bf VI versus MCMC for different $N$ values}. Clustering performance at $K=7$ across $N\in\{50, 200, 500, 1000\}$. We compare VI with a prior on $\alpha$ and MCMC with a prior on $\alpha$. We report the inferred number of clusters, inference and prediction runtimes, and train/test ARI and completeness. VI is substantially faster than MCMC while maintaining competitive clustering quality and high completeness on both the training and test cohorts.}
\label{tab:k7_alpha_comparison_compact}
\begin{tabular}{lcccccccc}
\toprule
Method & $N$ & Inf. Clust. & Time$_{\text{inf.}}$ (s) & Time$_{\text{pred.}}$ (s) & ARI$_{\text{train}}$ & ARI$_{\text{test}}$ & Compl.$_{\text{train}}$ & Compl.$_{\text{test}}$ \\
\midrule
VI    & 1000 & 6 & \textbf{2.44}     & \textbf{0.00} 
& \textbf{0.7364} & \textbf{0.6925} & \textbf{1.0000} & \textbf{1.0000} \\
MCMC & 1000 & 5 & 11668.16           & 0.42 
& 0.6318          & 0.6218          & \textbf{1.0000} & \textbf{1.0000} \\
\midrule
VI   & 500  & 6 & \textbf{3.03}     & \textbf{0.00}
& 0.4697          & 0.3694          & \textbf{0.9860} & \textbf{1.0000} \\
MCMC & 500  & 7 & 7617.04            & 0.42 
& \textbf{0.6703} & \textbf{0.6087} & 0.9721          & \textbf{1.0000} \\
\midrule
VI    & 200  & 6 & \textbf{0.99}     & \textbf{0.00} 
& 0.4167          & 0.1333          & 0.9433          & \textbf{1.0000} \\
MCMC  & 200  & 4 & 4466.42            & 0.43 
& \textbf{0.4733} & \textbf{0.2723} & \textbf{0.9690} & \textbf{1.0000} \\
\midrule
VI   & 50   & 5 & \textbf{0.42}     & \textbf{0.00}  
& \textbf{0.3953} & \textbf{0.3005} & \textbf{1.0000} & \textbf{1.0000} \\
MCMC & 50   & 5 & 2865.92            & 0.54 
& 0.3618          & 0.1839          & 0.9456          & 0.8727          \\
\bottomrule
\end{tabular}
\end{table*}

We further investigated the performance of the two inference approaches for $N = 200$ and $K = 7$, by creating 10 independent random train/test splits, where their performance is evaluated on each one of the splits. Both the inference and the prediction times of VI are substantially lower than the corresponding MCMC ones (Table~\ref{tab:vi_mcmc_runtime_repeats}). MCMC attains slighly higher mean ARI in both the training and test sets but the overlapping standard deviations of VI and MCMC suggest no major differences between these two approaches. Similar trends are also observed for NMI and homegeneity (Appendix Tables~\ref{tab:vi_mcmc_train_comparison} and~\ref{tab:vi_mcmc_test_comparison}). The recorded completeness of the VI approximation on the test set further supports its use as a decision-making tool (Table~\ref{tab:vi_mcmc_quality_repeats}).

\begin{table*}[!h]
\centering
\small
\caption{\textbf{VI versus MCMC over repeats}. Runtime and number of inferred clusters at $K=7$ and $N=200$, averaged over 10 independent random training/test set splits. The mean $\pm$ standard deviation for the number of inferred clusters and runtime are reported. VI is substantially faster than MCMC at both inference and prediction time.}
\label{tab:vi_mcmc_runtime_repeats}
\begin{tabular}{lccc}
\toprule
Method & Inf. Clust. & Inf. Time (s) & Pred. Time (s) \\
\midrule
VI   & $6.00 \pm 0.00$  & $\mathbf{1.00 \pm 0.22}$       & $\mathbf{0.00 \pm 0.00}$ \\
MCMC & $5.60 \pm 1.56$ & $5287.62 \pm 1580.97$          & $0.33 \pm 0.04$ \\
\bottomrule
\end{tabular}
\end{table*}

\begin{table*}[!h]
\centering
\small
\caption{\textbf{VI versus MCMC over repeats}. Clustering performance at $K=7$ and $N=200$, averaged over 10 independent random training/test set splits. The mean $\pm$ standard deviation for training/test sets ARI and completeness are reported. VI maintains high completeness on both cohorts, while MCMC attains slightly higher mean ARI, however the standard deviations overlap.}
\label{tab:vi_mcmc_quality_repeats}
\begin{tabular}{lcccc}
\toprule
Method & ARI$_{\text{train}}$ & ARI$_{\text{test}}$ & Compl.$_{\text{train}}$ & Compl.$_{\text{test}}$ \\
\midrule
VI   & $0.4119 \pm 0.0043$ & $0.4389 \pm 0.1817$ & $0.9611 \pm 0.0138$ & $\mathbf{1.0000 \pm 0.0000}$ \\
MCMC & $\mathbf{0.5327 \pm 0.1132}$ & $\mathbf{0.4771 \pm 0.2337}$ & $\mathbf{0.9756 \pm 0.0231}$ & $0.9627 \pm 0.0585$ \\
\bottomrule
\end{tabular}
\end{table*}

\begin{figure}[!h]
    \centering
    \begin{subfigure}[t]{\linewidth}
        \centering
        \includegraphics[width=0.875\linewidth]{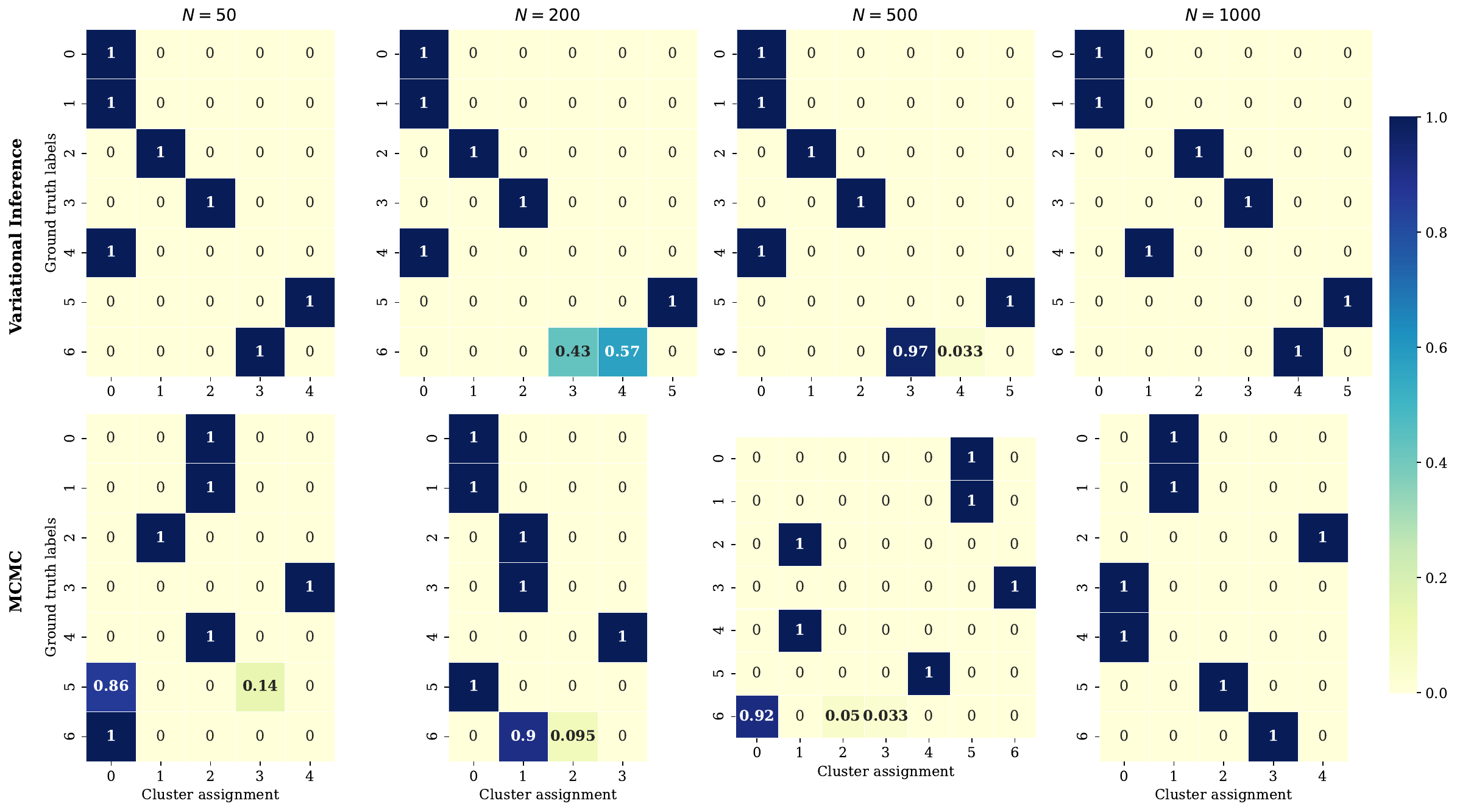}
        \caption{Training set.}
        \label{fig:heatmap_train_norm}
    \end{subfigure}
    
    \vspace{0.5em}
    
    \begin{subfigure}[t]{\linewidth}
        \centering
        \includegraphics[width=0.875\linewidth]{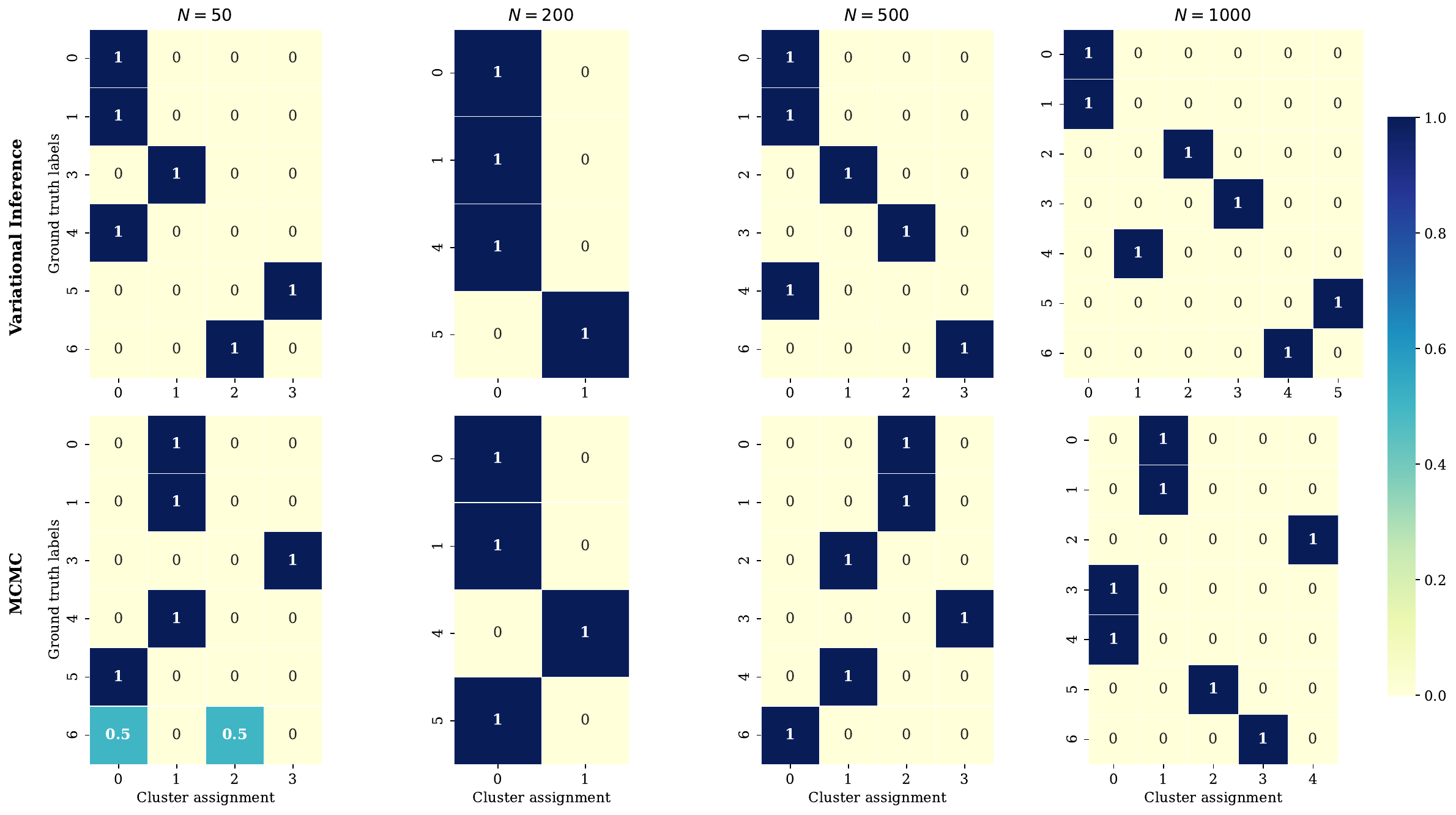}
        \caption{Test set.}
        \label{fig:heatmap_test_norm}
    \end{subfigure}
    
    \caption{{\bf Normalised contingency tables for VI and MCMC}. Comparison of VI and MCMC at $K = 7$ for $N\in\{50, 200, 500, 1000\}$. Each panel shows a contingency matrix with cluster assignments as columns and true classes as rows. Counts are normalised within each row to highlight whether observations from the same true class are assigned to the same predicted cluster. Despite most models predicting fewer clusters than the true classes, they generally group observations from the same true classes together.} 
    \label{fig:heatmap_norm}
\end{figure}

To further assess clustering quality, we examine contingency matrices that evaluate how well the cluster assignments recover the true underlying structure. We present in Figure~\ref{fig:heatmap_train_norm} heatmaps of the row normalised contingency matrix for the training set and in Figure~\ref{fig:heatmap_test_norm} for the test set. Each row corresponds to a ground-truth label and each column to an inferred cluster. On the training set both methods split true cluster 6 into two inferred clusters, for $N = 200$ and $N = 500$, and additionally MCMC did so on true cluster 5 for $N = 50$. On the test set, variational inference perfectly contains together true clusters in the inferred clusters, while MCMC splits true cluster 6 for $N = 50$. The corresponding unnormalised contingency matrices are presented in Appendix Figures~\ref{fig:heatmap_test} and~\ref{fig:heatmap_train} for the test and training sets, respectively. These visualisations further emphasise the accurate clustering offered by the VI approximation and are a tool that can be used for practitioners when they apply this approach. 

\section{Discussion} \label{sec:discussion}
We propose a VI approximation for DPGMM as a scalable and efficient clustering tool. Through a comprehensive numerical study we illustrated that the VI implementation is an effective clustering tool achieving strong performance as measured by multiple clustering metrics. The VI implementation was found to have similar, or even better, performance to its counterpart MCMC with a significantly reduced computational running time.  We acknowledge that the simulated data limits the ability to fully capture the complexity of real biomedical settings. Additionally, we acknowledge deviations from model assumptions, may affect model performance. An extension would be to evaluate the approach on real datasets with clinically validated subtype labels, for example those curated or annotated by medical professionals. Such work will further enable the proposed VI approximation for DPGMM to be used as part of a medical decision-making framework.

\begin{ack}
This work was supported by the Engineering and Physical Sciences Research Council (EPSRC)
through the Modern Statistics and Statistical Machine Learning (StatML) CDT programme, grant no. EP/Y034813/1 and CRUK Convergence Science Centre ref. CANCTA-2026/100007. 
\end{ack}

\bibliography{references}
\bibliographystyle{plainnat}


\appendix

\section{ELBO}\label{sec:elbo}

In this section we provide the explicit form of the evidence lower bound (ELBO) for the proposed DPGMM. For the model defined in Equation~\ref{eq:DPMM_2}, let the latent parameters be $\boldsymbol{w} = \{\boldsymbol{\beta}, \boldsymbol{\mu}^*, \boldsymbol{\sigma}^2{^*}, \boldsymbol{z}, \alpha\}$ and the hyperparameters be $\{\mu_0, \sigma^2_0, b, c, s_1, s_2\}$. The ELBO in Equation~\ref{eq:ELBO} can then be written as follows,

\begin{align*}\label{eq:ELBO_DPMM}
    \log p(\boldsymbol{x}|\mu_0, \sigma^2_0, b, c, s_1, s_2) &\geq E_q[\log p(\boldsymbol{\beta}|\alpha)]  +
    E_q[\log p(\alpha|s_1, s_2)] \nonumber \\ &+ E_q[\log p(\boldsymbol{\mu}^*|\mu_0, \sigma^2_0)]  + 
    E_q[\log p(\boldsymbol{\sigma}^2{^*}|b, c)] \nonumber \\
   & +\sum^N_{n=1}\left(E_q[\log p(Z_n| \boldsymbol{\beta})] + E_q[\log p(x_{n}|Z_n,\boldsymbol{\mu}{^*},\boldsymbol{\sigma}^2{^*})]\right)
   \nonumber \\
   &- E_q[\log q(\boldsymbol{\beta}, \alpha, \boldsymbol{\mu}{^*},\boldsymbol{\sigma}^2{^*}, \boldsymbol{z})].
\end{align*}

The expectation of the  log joint distribution under the variational approximation,
$E_q[ \log p(\boldsymbol{w}, \boldsymbol{x}) ]$ is given by,

\begin{align*}
   E_q[ \log p(\boldsymbol{w}, \boldsymbol{x}) ] \propto &\sum^\infty_{k=1} 
   \psi(w_1) - \log(w_2) + \left(\frac{w_1}{w_2}-1\right)(\psi(\rho_{k,2})-\psi(\rho_{k,2}+\rho_{k,1}))\\
   &+(s_1-1)(\psi(w_1)-\log(w_2))-s_2\frac{w_1}{w_2}\\
    & -\frac{1}{2}\sum^\infty_{k=1} (\tau_{k,1}^{\mu}-\mu_0 )^T\left(\frac{1}{\sigma^2_0}\boldsymbol{I}\right)(\tau_{k,1}^{\mu}-\mu_0 ) + \text{trace}\left(\frac{1}{\sigma^2_0}\boldsymbol{I}\tau_{k,2}^{\mu}\right)
    \\
      &+ (-b-1)(\log(\tau_{k,2}^{\sigma^2})-\psi(\tau_{k,1}^{\sigma^2})-c\frac{\tau_{k,1}^{\sigma^2}}{\tau_{k,2}^{\sigma^2}}\\
    &+ \sum^N_{n=1} \sum^\infty_{k=1} \left(\sum^\infty_{l=k+1}\phi_{n,l}(\psi(\rho_{k,2})+\psi(\rho_{k,2}+\rho_{k,1}))\right) +\phi_{n,k}(\psi(\rho_{k,1})+\psi(\rho_{k,2}+\rho_{k,1}))\\
    &+\phi_{n,k} \left[ P\log (2\pi)+P(\log(\tau_{k,2}^{\sigma^2})-\psi(\tau_{k,1}^{\sigma^2})) + \frac{\tau_{k,1}^{\sigma^2}}{\tau_{k,2}^{\sigma^2}}\left(||x_{n,j}-\tau^\mu_{k,1}||^2_2 + \text{trace}(\tau^\mu_{k,2}) \right)\right].
\end{align*}

The expectation of the log of variational approximation,  $E_q[\log q(\boldsymbol{w})]$, is the following,

\begin{align*}
    E_q[\log q(\boldsymbol{w})] &\propto  
      \sum^\infty_{k=1} \log \Gamma(\rho_{k,1}+\rho_{k,2})-\log\Gamma (\rho_{k,1})- \log \Gamma(\rho_{k,2})\\&+ (\rho_{k,1}-1) (\psi(\rho_{k,1})- \psi(\rho_{k,1}+\rho_{k,2})) \\
    &+ (\rho_{k,2}-1) (\psi(\rho_{k,2})- \psi(\rho_{k,1}+\rho_{k,2}))  \\
    &+(w_1-1)(\psi(w_1)-\log(w_2))-w_1+w_1\log(w_2)-\log(\Gamma(w_1))\\
    &-\frac{1}{2} \sum^\infty_{k=1} \log \det \tau^{\mu}_{k,2} \\
    &+\sum^\infty_{k=1}(\tau^{\sigma^2}_{k,1}+1)\psi(\tau^{\sigma^2}_{k,1})-\log(\tau^{\sigma^2}_{k,2})-\tau^{\sigma^2}_{k,1}-\log(\Gamma(\tau^{\sigma^2}_{k,2}))\\
    &+ \phi_{n,k}\log(\phi_{n,k}).
\end{align*}

\section{Numerical Study Setup}\label{sec:numerical_study_setup}

In this section, we provide additional implementation details to complement Section~\ref{sec:numerical_studies} and support reproducibility. All experiments were implemented in \texttt{Python} and were run on a standard laptop CPU (MacBook Pro with Apple M3 chip and 8GB RAM) to better reflect practical settings relevant to healthcare practitioners. All runtimes are reported as wall-clock runtimes, and all performance evaluation metrics were computed using \texttt{scikit-learn} \citep{scikit-learn}. The corresponding \texttt{Python} code for data generation, implementation of the proposed CAVI algorithm, and the MCMC comparison implementation in \texttt{PyMC} is provided in \url{https://github.com/ingahuld/DPMM}. 

\subsection{Data Generation}

We present the data generating process in detail. We set $K\in\{2,...,11\}$, $N\in\{60,210,510,1100\}$ as for each we sample both training data and extra 10 observations for test set, $P = 10$. 

\begin{enumerate}
    \item Set seed to 308. 
    \item For each mixture, $k\in \{1,...,K\}$, sample the corresponding mean, variance and mixture probability.
    \begin{align*}
        \pi_k &\sim \text{Dirichlet}(10\cdot \boldsymbol{1}_k)\\
        \mu_k &\sim \mathcal{N}(0,5)\\
        \sigma^2_k &\sim \text{Inverse-Gamma}(2,1)
    \end{align*}
    \item Sample each observation $i\in \{1,...,N\}$ from a Gaussian with relevant parameters sampled in the previous step. Use the variance to define the autoregressive correlation matrix.  
    \begin{align*}
        Y_i = k &\sim \text{Categorical}(\pi_1,...,\pi_K)\\
        \Sigma_{ij}^{(k)} &= \sigma^2_k\rho^{|i-j|}, \quad \rho = 0.3, \quad j\in\{1,...,P\}\\
        X_i &\sim \mathcal{N}_P(\mu_k\boldsymbol{1}_P, \Sigma^{(k)})
    \end{align*}
    \item Introduce additive Gaussian noise to each observation $X_{ij}$, where $i\in \{1,...,N\}$ and $j\in\{1,...,P\}$.
    \begin{align*}
    \sigma_{\epsilon}^2 &\sim \Gamma(0.5, 0.1),\\
        \epsilon_{ij} &\sim \mathcal{N}(0,\sigma^2_{\epsilon})\\
        X_{ij}  &= X_{ij} + \epsilon_{ij}
    \end{align*}
    \item Introduce noisy features to each observation $n\in \{1,...,N\}$
    \begin{align*}
        P_{\text{noise}} &= \lfloor p\cdot P \rfloor \quad p = 0.25\\
        P_{\text{signal}} &= P - P_{\text{noise}}\\
        X_{ij_\text{noise}} &\sim \mathcal{N}(0,1), \quad j_\text{noise}\in\{1,...,P_{\text{noise}}\}
    \end{align*}
    \begin{itemize}
        \item Keep first $P_{\text{signal}}$ columns of each observation and concatenate the $P_{\text{signal}}$ columns from $X_{ij_\text{noise}}$ and permute all columns. 
    \end{itemize}
    \item Final product: 
    \begin{align*}
        X \in \mathbb{R}^{N\times P}, Y\in \mathbb{N}^{N}
    \end{align*}
\end{enumerate}

When generating the train test split or multiple train test splits for repeats, the random state is set as 42. 

\subsection{Variational Inference}

The update steps of the CAVI algorithm are implemented corresponding to the equations in Section~\ref{sec:model}. To perform predictions on new data observations, the update step for $\phi_n$ is run given the other parameter values and the new data observations. We monitor convergence of the algorithm by tracking the ELBO values and set the covergence criteria to $10^{-5}$. The truncation parameter of the model is set to $T = 20$.

\subsubsection{Initialisation}

The variational parameters are initialised as follows,

\begin{itemize}
    \item $\phi_n$: equal probability over each possible cluster assignment.
    \item $\tau^{\mu}_k$: All set to zero.
    \item $\tau^{\sigma^2}_k$: All set to one.
    \item $\rho_1$: All set to one. 
    \item $\rho_2$: All set to $\alpha$. If there is a $\text{Gamma}(s_1,s_2)$ prior on $\alpha$, all set to E$[\alpha] = \frac{s_1}{s_2}$. 
    \item $w_1$, $w_2$: Set to $s_1$, $s_2$ respectively. 
\end{itemize}

\subsection{MCMC}

To implement the MCMC inference approach of the DPGMM we use the open source package \texttt{PyMC} \citep{abril2023pymc} with the \texttt{numpyro} sampler that leverages a \texttt{JAX} backend. We sample 1,000 samples and 500 burning samples using 4 cores. As PyMC does not have the functionality of sampling discrete variables, such as cluster assignments, their approach is to marginalise out the parameter. As demonstrated by \citet{pymctutorial} the cluster assignments are implemented by computing the posterior probability,

\begin{align*}
    \frac{\pi_k\cdot p(z_i = k|x,\mu_k, \sigma^2_k)}{\sum_k \pi_k\cdot p(z_i = k|x, \mu_k, \sigma^2_k)},
\end{align*}

for each possible cluster for the new data over all the samples and averaging over the samples, the cluster assignment is made by taking the maximum value over each posterior responsibility. The random state is set as 307, and for repeats it is set as $307+i$ where $i$ is the current iteration number.

\clearpage
\newpage
\section{Additional results}\label{sec:appendix_results}

\begin{table*}[h]
\centering
\small
\caption{VI training set clustering performance for increasing values of $K$ and decreasing dataset size $N$ within each block. A Gamma$(1,1)$ prior is assumed for $\alpha$. The inferred number of clusters, inference runtime, training set ARI, NMI, homogeneity, completeness, and silhouette score are reported.}
\label{tab:vi_alpha_prior_train}
\begin{tabular}{ccccccccc}
\toprule
$K$ & $N$ & Clust. & Inf. Time & ARI$_{\text{train}}$ & NMI$_{\text{train}}$ & Homog.$_{\text{train}}$ & Compl.$_{\text{train}}$ & Silhouette \\
\midrule
2 & 1000 & 2 & 2.83  & 1.0000 & 1.0000 & 1.0000 & 1.0000 & 0.8733 \\
2 & 500  & 2 & 1.57  & 1.0000 & 1.0000 & 1.0000 & 1.0000 & 0.8730 \\
2 & 200  & 3 & 2.11  & 0.8425 & 0.8519 & 1.0000 & 0.7420 & 0.4119 \\
2 & 50   & 2 & 0.47  & 1.0000 & 1.0000 & 1.0000 & 1.0000 & 0.8711 \\
\midrule
3 & 1000 & 5 & 5.94  & 0.9905 & 0.9773 & 1.0000 & 0.9556 & 0.3584 \\
3 & 500  & 4 & 43.20 & 0.8274 & 0.8864 & 1.0000 & 0.7960 & 0.4439 \\
3 & 200  & 3 & 0.73  & 1.0000 & 1.0000 & 1.0000 & 1.0000 & 0.4557 \\
3 & 50   & 3 & 0.65  & 1.0000 & 1.0000 & 1.0000 & 1.0000 & 0.4188 \\
\midrule
7 & 1000 & 6 & 2.44  & 0.7364 & 0.9170 & 0.8468 & 1.0000 & 0.4914 \\
7 & 500  & 6 & 3.03  & 0.4697 & 0.7927 & 0.6627 & 0.9860 & 0.4075 \\
7 & 200  & 6 & 0.99  & 0.4167 & 0.7634 & 0.6411 & 0.9433 & 0.3829 \\
7 & 50   & 5 & 0.42  & 0.3953 & 0.7611 & 0.6143 & 1.0000 & 0.3983 \\
\bottomrule
\end{tabular}
\end{table*}

\begin{table*}[!h]
\centering
\small
\caption{VI test-set clustering performance across increasing values of $K$ and decreasing dataset size $N$ within each block. A Gamma$(1,1)$ prior is assumed for $\alpha$. The prediction runtime, test set ARI, NMI, homogeneity, and completeness are reported.}
\label{tab:vi_alpha_prior_test}
\begin{tabular}{ccccccc}
\toprule
$K$ & $N$ & Pred. Time & ARI$_{\text{test}}$ & NMI$_{\text{test}}$ & Homog.$_{\text{test}}$ & Compl.$_{\text{test}}$ \\
\midrule
2 & 1000 & 0.00039 & 1.0000 & 1.0000 & 1.0000 & 1.0000 \\
2 & 500  & 0.00036 & 1.0000 & 1.0000 & 1.0000 & 1.0000 \\
2 & 200  & 0.00038 & 1.0000 & 1.0000 & 1.0000 & 1.0000 \\
2 & 50   & 0.00037 & 1.0000 & 1.0000 & 1.0000 & 1.0000 \\
\midrule
3 & 1000 & 0.00040 & 1.0000 & 1.0000 & 1.0000 & 1.0000 \\
3 & 500  & 0.00049 & 1.0000 & 1.0000 & 1.0000 & 1.0000 \\
3 & 200  & 0.00187 & 1.0000 & 1.0000 & 1.0000 & 1.0000 \\
3 & 50   & 0.00045 & 1.0000 & 1.0000 & 1.0000 & 1.0000 \\
\midrule
7 & 1000 & 0.00038 & 0.6925 & 0.9347 & 0.8774 & 1.0000 \\
7 & 500  & 0.00121 & 0.3694 & 0.7821 & 0.6421 & 1.0000 \\
7 & 200  & 0.00144 & 0.1333 & 0.4051 & 0.2540 & 1.0000 \\
7 & 50   & 0.00037 & 0.3005 & 0.7677 & 0.6229 & 1.0000 \\
\bottomrule
\end{tabular}
\end{table*}

\begin{figure}[!h]
  \centering
      \includegraphics[width=0.95\linewidth]{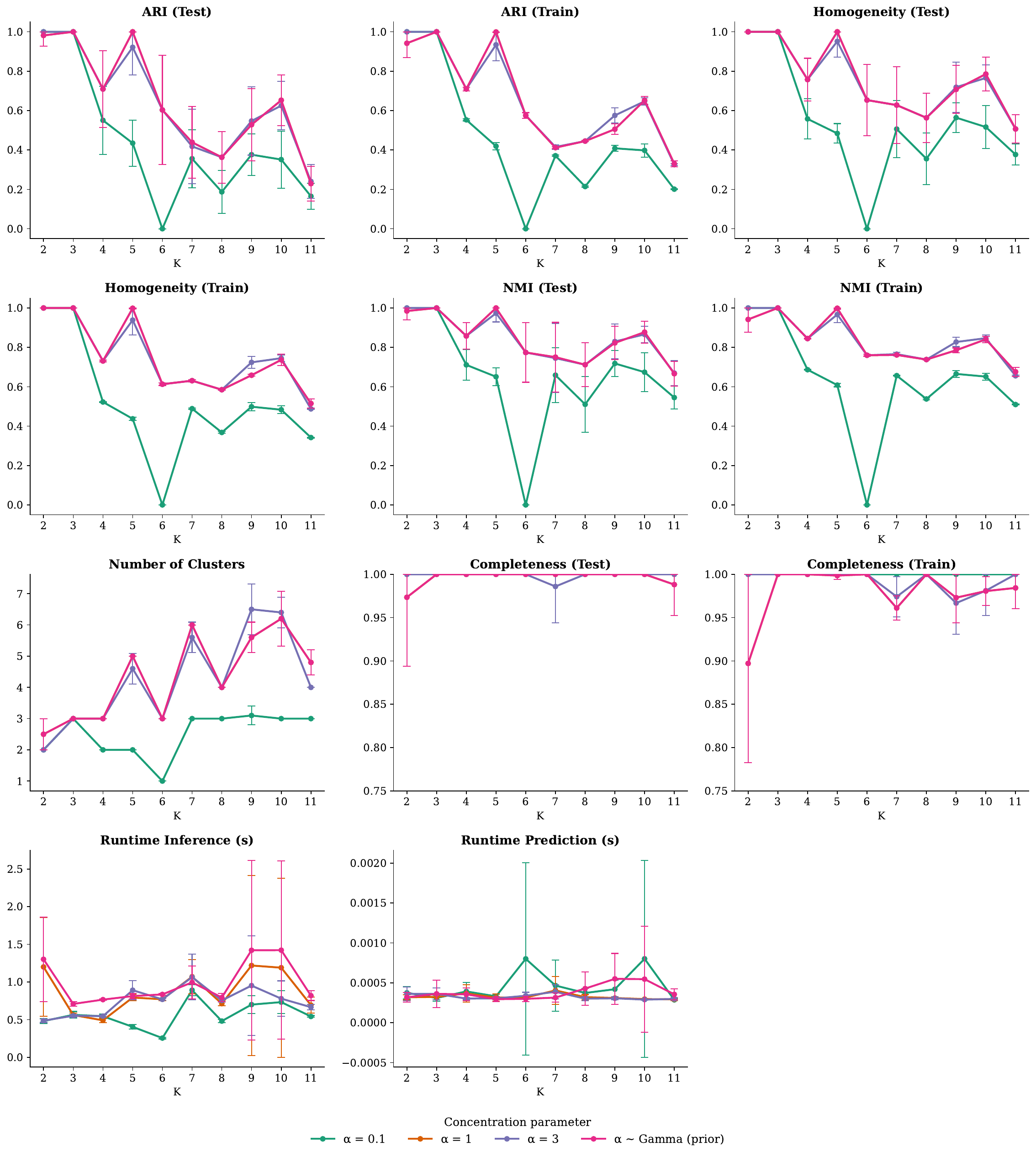}
  \caption{VI clustering performance over 10 independent random training/test splits for $K\in \{2, 3,..., 9, 11\}$, $\alpha \in \{0.1, 1, 3\}$ and $\alpha \sim \text{Gamma}(1,1)$, and $N = 200$. The mean, and standard deviation of ARI and NMI, homogeneity, completeness, number of inferred clusters, and runtime on both training and test data are illustrated.}
  \label{fig:vi_repeats_metrics_appendix}
\end{figure}

\begin{figure}[!h]
  \centering
      \includegraphics[width=0.95\linewidth]{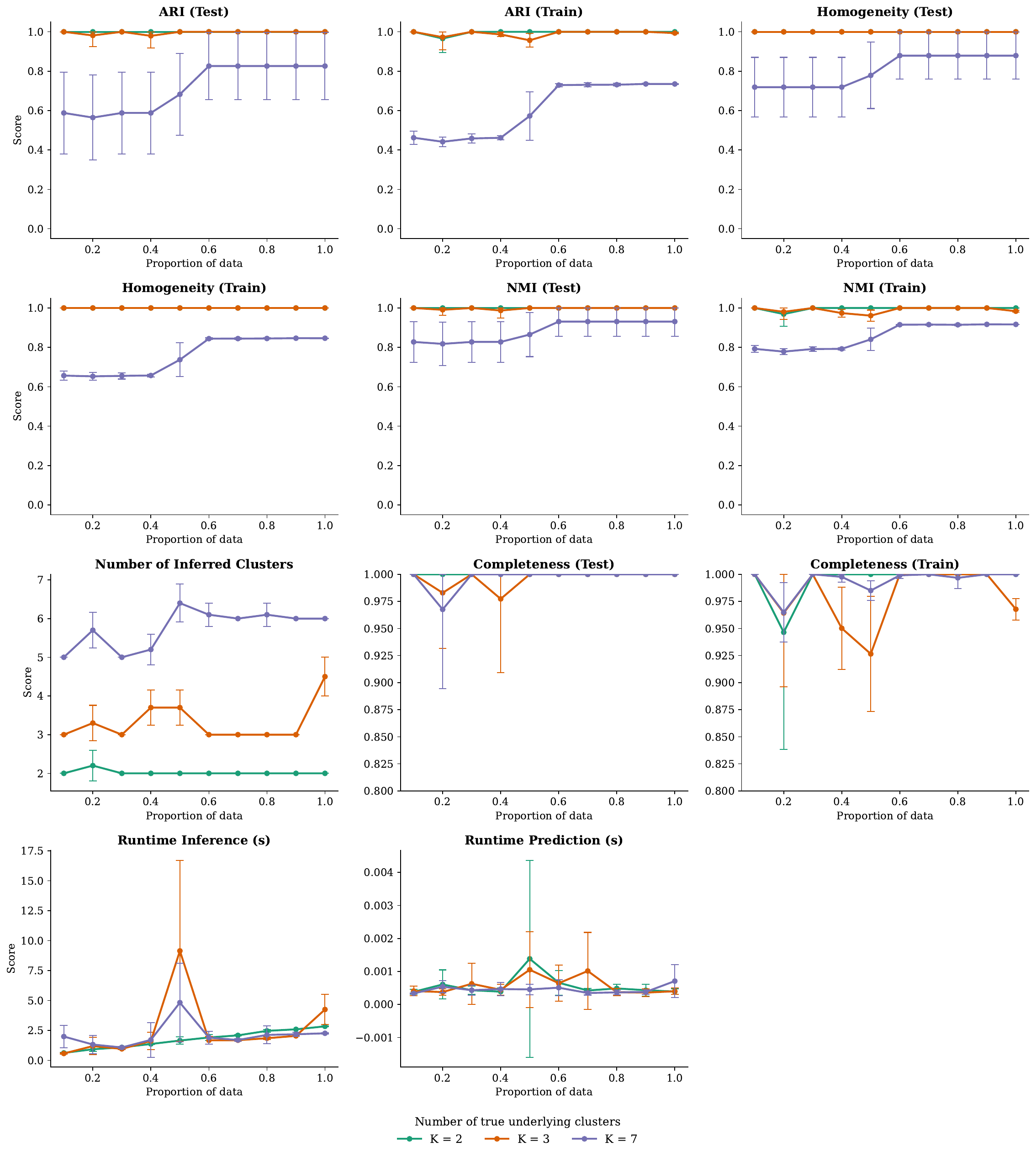}
  \caption{VI clustering performance as the number of training observations increases, with $N_{\text{train}} = 1000p$ for $p\in\{0.1,0.2,...,0.9,1.0\}$. The mean, and standard deviation over 10 train/test splits are illustrated for inference runtime, test set ARI, training and test sets completeness. The mean, and standard deviation of ARI and NMI, homogeneity, completeness, number of inferred clusters, and runtime on both training and test data are illustrated.}
  \label{fig:vi_repeats_metrics_increaseN_appendix}
\end{figure}

\clearpage
\newpage 

\begin{table*}[!h]
\centering
\small
\caption{Training set clustering performance for VI and MCMC at $K=7$ across decreasing dataset sizes $N$. A prior Gamma$(1,1)$ is assumed for $\alpha$. The inferred number of clusters, inference runtime, training set ARI, NMI, homogeneity, completeness, and silhouette score are reported.}
\label{tab:k7_alpha_comparison_train}
\begin{tabular}{lcccccccc}
\toprule
Method & $N$ & Inf. Clust. & Inf. Time & ARI$_{\text{train}}$ & NMI$_{\text{train}}$ & Homog.$_{\text{train}}$ & Compl.$_{\text{train}}$ & Silhouette \\
\midrule
VI       & 1000 & 6 & \textbf{2.44}     & \textbf{0.7364} & \textbf{0.9170} & \textbf{0.8468} & \textbf{1.0000} & 0.4914 \\
MCMC    & 1000 & 5 & 11668.16 & 0.6318 & 0.8474 & 0.7352 & \textbf{1.0000} & 0.2883 \\
\midrule
VI     & 500  & 6 & \textbf{3.03}     & 0.4697 & 0.7927 & 0.6627 & \textbf{0.9860} & 0.4075 \\
MCMC      & 500  & 7 & 7617.04  & \textbf{0.6703} & \textbf{0.8571} & \textbf{0.7664} & 0.9721 & 0.4698 \\
\midrule
VI      & 200  & 6 & \textbf{0.99}     & 0.4167 & \textbf{0.7634} & \textbf{0.6411} & \textbf{0.9433} & 0.3829 \\
MCMC     & 200  & 4 & 4466.42  & \textbf{0.4733} & 0.7046 & 0.5536 & 0.9690 & 0.0907 \\

\midrule
VI    & 50   & 5 & \textbf{0.42}     & \textbf{0.3953} & \textbf{0.7611} & \textbf{0.6143} & \textbf{1.0000} & 0.3983 \\
MCMC   & 50   & 5 & 2865.92  & 0.3618 & 0.6963 & 0.5511 & 0.9456 & 0.3343 \\

\bottomrule
\end{tabular}
\end{table*}

\begin{table*}[!h]
\centering
\small
\caption{Test set clustering performance for VI and MCMC at $K=7$ across decreasing dataset sizes $N$. A prior Gamma$(1,1)$ is assumed for $\alpha$. The prediction runtime, test set ARI, NMI, homogeneity, and completeness are reported.} 
\label{tab:k7_alpha_comparison_test}
\begin{tabular}{lcccccc}
\toprule
Method & $N$ & Pred. Time & ARI$_{\text{test}}$ & NMI$_{\text{test}}$ & Homog.$_{\text{test}}$ & Compl.$_{\text{test}}$ \\
\midrule
VI     & 1000 & \textbf{0.00038} & \textbf{0.6925} & \textbf{0.9347} & \textbf{0.8774} & \textbf{1.0000} \\
MCMC   & 1000 & 0.4196           & 0.6218          & 0.8900          & 0.8018          & \textbf{1.0000} \\
\midrule
VI     & 500  & \textbf{0.00121} & 0.3694          & 0.7821          & 0.6421          & \textbf{1.0000} \\
MCMC   & 500  & 0.4225           & \textbf{0.6087} & \textbf{0.8602} & \textbf{0.7547} & \textbf{1.0000} \\

\midrule
VI      & 200  & \textbf{0.00144} & 0.1333          & 0.4051          & 0.2540          & \textbf{1.0000} \\
MCMC   & 200  & 0.4326           & \textbf{0.2723} & \textbf{0.5622} & \textbf{0.3910} & \textbf{1.0000} \\

\midrule
VI        & 50   & \textbf{0.00037} & \textbf{0.3005}          & \textbf{0.7677}          & \textbf{0.6229}          & \textbf{1.0000} \\
MCMC      & 50   & 0.5386           & 0.1839          & 0.6699          & 0.5436          & 0.8727          \\

\bottomrule
\end{tabular}
\end{table*}

\begin{sidewaystable}[h]
\centering
\small
\caption{Comparison of VI and MCMC at $K=7$ and $N=200$, averaged over 10 independent random training/test splits. The mean $\pm$ standard deviation for the inferred number of clusters, inference runtime, training set ARI, NMI, homogeneity, completeness, and silhouette score are reported.}
\label{tab:vi_mcmc_train_comparison}
\begin{tabular}{lccccccc}
\toprule
Method & Clust. & Inf. Time & ARI$_{\text{train}}$ & NMI$_{\text{train}}$ & Homog.$_{\text{train}}$ & Compl.$_{\text{train}}$ & Silhouette \\
\midrule
VI   & $6.0 \pm 0.0$  & $1.00 \pm 0.22$       & $0.4119 \pm 0.0043$ & $0.7616 \pm 0.0033$ & $0.6308 \pm 0.0057$ & $0.9611 \pm 0.0138$ & $0.3901 \pm 0.0081$ \\
MCMC & $5.6 \pm 1.56$ & $5287.62 \pm 1580.97$ & $0.5327 \pm 0.1132$ & $0.7795 \pm 0.0743$ & $0.6558 \pm 0.0999$ & $0.9756 \pm 0.0231$ & $0.3327 \pm 0.1234$ \\
\bottomrule
\end{tabular}
\end{sidewaystable}

\begin{table}[htbp]
\centering
\small
\caption{Comparison of VI and MCMC at $K=7$ and $N=200$, averaged over 10 independent random train/test splits. The mean $\pm$ standard deviation for prediction runtime and test set ARI, NMI, homogeneity, and completeness are reported.}
\label{tab:vi_mcmc_test_comparison}
\begin{tabular}{lccccc}
\toprule
Method & Pred. Time & ARI$_{\text{test}}$ & NMI$_{\text{test}}$ & Homog.$_{\text{test}}$ & Compl.$_{\text{test}}$ \\
\midrule
VI   & $0.000315 \pm 0.000057$ & $0.4389 \pm 0.1817$ & $0.7503 \pm 0.1766$ & $0.6278 \pm 0.1951$ & $1.0000 \pm 0.0000$ \\
MCMC & $0.3272 \pm 0.0361$     & $0.4771 \pm 0.2337$ & $0.7669 \pm 0.1279$ & $0.6576 \pm 0.1750$ & $0.9627 \pm 0.0585$ \\
\bottomrule
\end{tabular}
\end{table}

\clearpage
\newpage 

\begin{figure}[!h]
  \centering
      \includegraphics[width=0.95\linewidth]{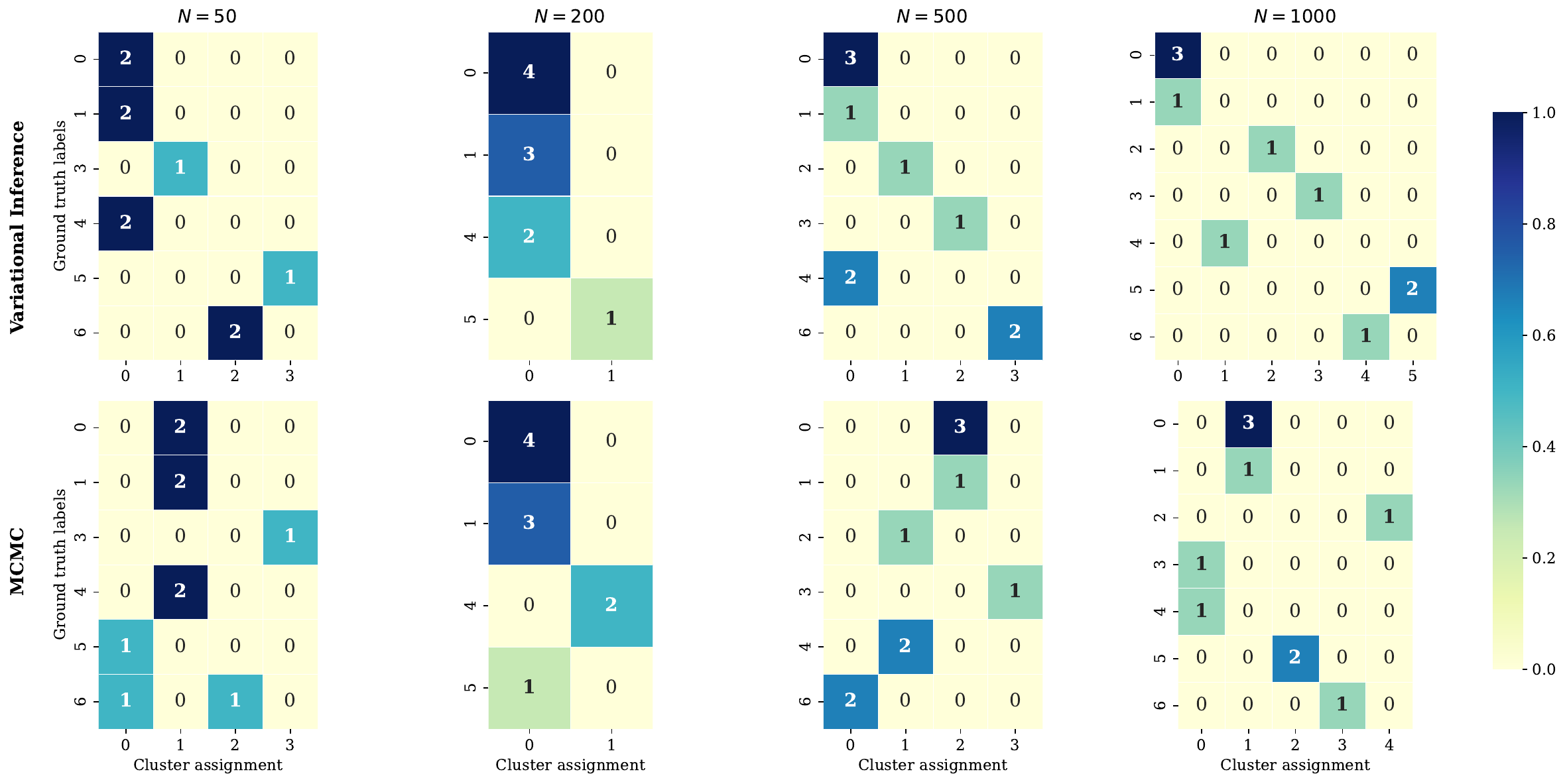}
  \caption{Unnormalised contingency matrices for the VI and MCMC cluster assignments at $K = 7$ for $N \in \{50, 200, 500, 1000\}$. The cluster assignments are computed on a held out dataset of 10 test observations. The contingency matrix, containing frequency counts of test observations distributed over the predicted clusters by the columns and the true classes by rows.}
  \label{fig:heatmap_test}
\end{figure}

\begin{figure}[!h]
  \centering
      \includegraphics[width=0.95\linewidth]{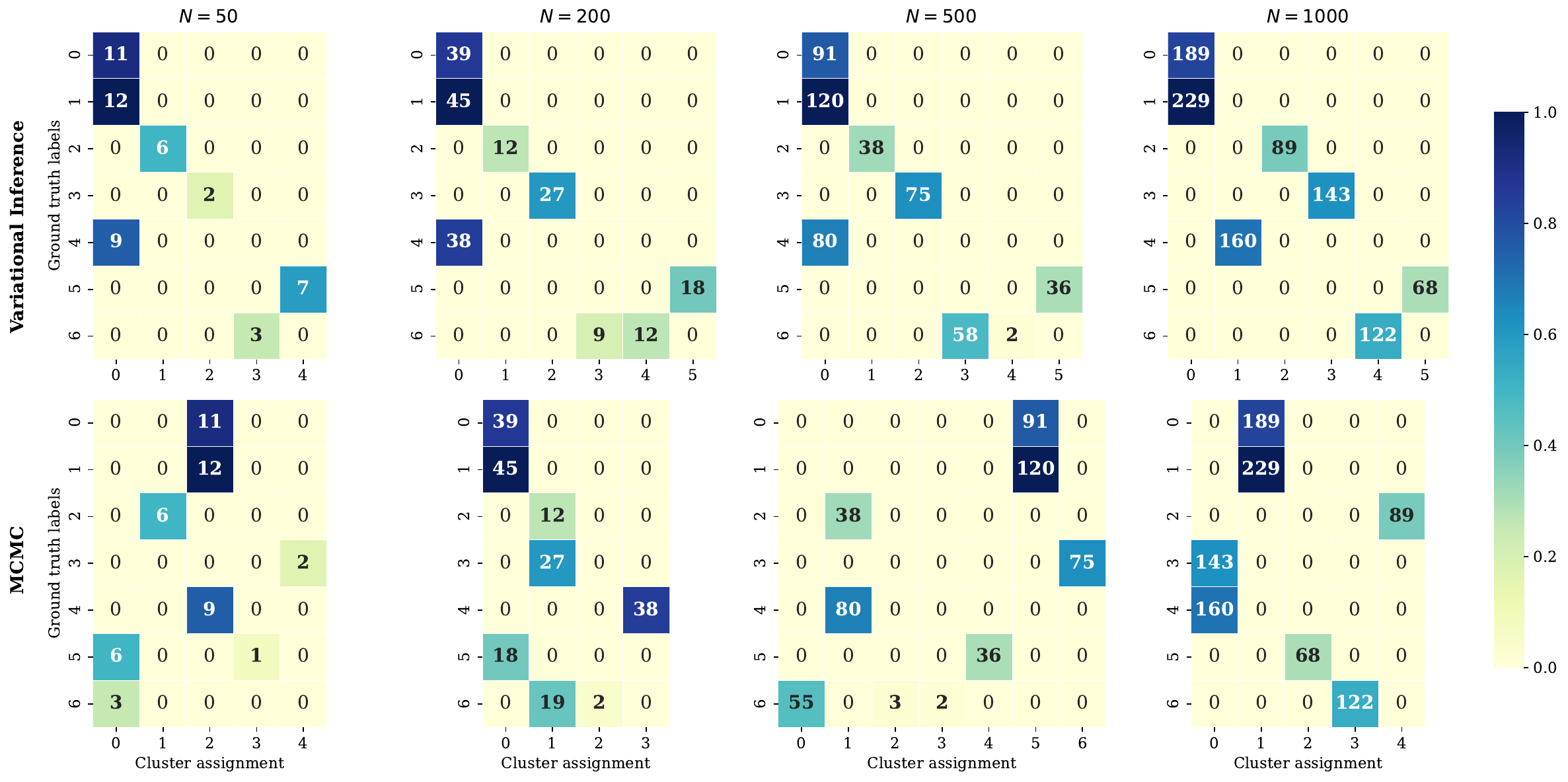}
  \caption{Unnormalised contingency matrices for the VI and MCMC cluster assignments at $K = 7$ for $N \in \{50, 200, 500, 1000\}$. The cluster assignments are computed on the training set. The contingency matrix, containing frequency counts of the training set observations distributed over the predicted clusters by the columns and the true classes by rows. }
  \label{fig:heatmap_train}
\end{figure}



\end{document}